\documentclass[oldversion,rnote,preprint]{aa}  
\usepackage{graphicx}
\usepackage{txfonts}
\usepackage{longtable}
\usepackage{multirow}

\begin{document}

\title{Over-resolution of compact sources in interferometric observations}

   \author{I. Mart\'i-Vidal\inst{1,2}
          \and
          M.A. P\'erez-Torres\inst{3}
          \and
          A. P. Lobanov\inst{2}}
   \institute{
             Onsala Space Observatory,
             SE-439 92 Onsala (Sweden)
             \email{ivan.marti-vidal@chalmers.se}
             \and
             Max-Planck-Institut f\"ur Radioastronomie,
             Auf dem H\"ugel 69, D-53121 Bonn (Germany) 
             \and
             Instituto de Astrof\'isica de Andaluc\'ia, CSIC, 
             Apdo. Correos 2004, E-08071 Granada (Spain)}

   \date{Accepted for publication in A\&A}

\abstract{
We review the effects of source size in interferometric observations and focus on the 
cases of very compact sources.
If a source is extremely compact and/or weak (so it is not possible to detect signature of 
source structure in the visibilities) we describe a test of hypothesis that can be used to set 
a strong upper limit to the size of the source. We also estimate the minimum possible size of a source 
whose structure can still be detected by an interferometer (i.e., the maximum theoretical 
over-resolution power of an interferometer), which depends on the overall observing time, the 
compactness in the array distribution, and the sensitivity of the receivers. 
As a result, and depending on the observing frequency, the over-resolution power of 
forthcoming ultra-sensitive arrays, like the Square Kilometer Array (SKA), may allow us to study 
details of sources at angular scales down to a few $\mu$as.
}
   \maketitle

\section{Introduction}

Sensitivity and resolution (both spectral and angular) are the main limiting factors
in observational Astronomy. In the case of angular (i.e., spatial) resolution, 
the strong limitation that will always affect the observations, regardless of 
the quality of our instruments, is the {\em diffraction limit}. When an 
instrument 
is diffraction-limited, its response to a plane wave (i.e., to a point source 
located at infinity) is the so-called {\em Point Spread Function} (PSF), which 
has a width related to the smallest angular scale that can be resolved with the 
instrument.

It is well-known that the diffraction limit decreases with both an increasing 
observing frequency and an increasing aperture of the instrument. Hence, the only 
way to achieve a higher angular resolution at a given frequency is to increase the 
instrument aperture.
In this sense, the {\em aperture synthesis}, which is a technique related to 
astronomical interferometry (see, e.g., Thomson, Moran \& Swenson \cite{TMS}), 
presently seems to be almost the only way to further increase the angular 
resolution currently achieved at any wavelength.

But there is a crucial difference in interferometric observations, compared to 
those obtained with other techniques. When aperture synthesis is performed, 
an interferometer does not directly observe the structure of a source, but samples 
a fractions of its Fourier transform (the so-called {\em visibilities}). In other 
words, the space from which an interferometer takes measurements is not the sky 
itself, but the Fourier transform of its intensity distribution over 
the whole field of view. 
This special characteristic of interferometers strongly affects how these devices 
behave when we observe sources of sizes well below the diffraction limit, as 
we will see in the following sections.

It is possible, of course, to compute the inverse Fourier transform of a set of 
visibilities and (try to) recover the intensity distribution of the observed sources
in the sky. When combined with certain 
deconvolution algorithms, this approach of {\em imaging} a set of visibilities may 
be very useful if we are dealing with relatively extended sources. However, if the sources 
observed are very compact (relative to the diffraction limit of the interferometer), 
important and uncontrollable effects may arise in the imaging of the source intensity 
profiles, either coming from the (non-linear) deconvolution algorithms and/or from the 
gridding (pixelation) in the sky plane. These effects may result in strong biases 
in the estimate of source sizes, based in measurements performed in the sky plane.

The over-resolution power of an interferometer is a function of the baseline
sensitivity, and may play an important role in the analysis of data coming from
future ultra-sensitive interferometric arrays (like the square kilometer array, SKA,
or the Atacama large millimeter array, ALMA).

In this research note, we review several well-known aspects related to the 
effect of source compactness in visibility space, showing that it is possible 
to find out, from the observed visibilities, information on the size of sources 
much smaller than the diffraction limit achieved in the aperture synthesis. 
We also estimate the maximum theoretical over-resolution power of an 
interferometer and discuss a statistical test to estimate upper bounds to the size 
of ultra-compact sources observed with high-sensitivity interferometers.
In Lobanov (\cite{Lobanov}), the reader will find a detailed discussion of the 
resolution limits obtained for specific shapes of the brightness distribution. 
Here, we extend the discussion to a more general case of super-resolution that 
can be achieved with interferometers.


\section{Compact sources in visibility space}

The size of a source slightly smaller than the diffraction limit of an 
interferometer leaves a very clear fingerprint in visibility space, although 
its effect on the sky plane (after all the imaging and deconvolution 
steps) may be much less clear. 
For instance, if the size of a source is similar to the full width at half 
maximum (FWHM) of the PSF, the radial profile in the visibility amplitudes 
will decrease with baseline length falling down to $\sim$0.5 times the maximum 
visibility amplitude at the maximum projected baseline.
Hence, a source with a size of the order of the diffraction limit maps into an amplitude 
profile in visibility space that can be well detected and characterized by an 
interferometer.

If the size of the source decreases, the visibility amplitude in the 
longest baseline increases; in the limit case when the size of the source 
tends to zero, the visibility amplitude in the longest baseline tends to be 
as large as that in the shortest baseline. 
It is common, indeed, to compute the amplitude ratio between the visibilities in the 
shortest baseline and those in the longest baseline as a quantitative representation 
of the {\em degree of compactness} of the observed sources (e.g., Kovalev et al. 
\cite{Kovalev}; Lobanov \cite{Lobanov}). 

\begin{figure}
\centering
\includegraphics[width=8cm]{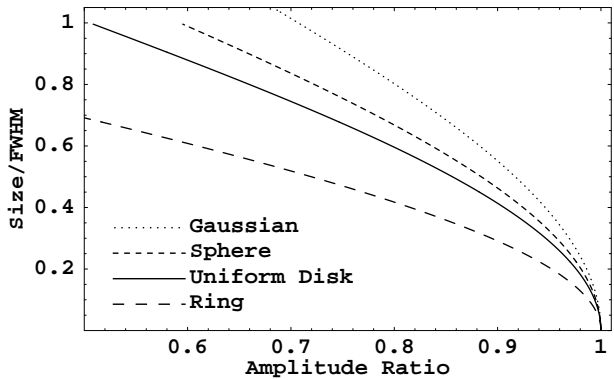}
\caption{Source size (in units of the FWHM of the synthesized beam) as 
a function of the ratio of the visibility amplitude in the longest 
baseline to that in the shortest baseline.}
\label{fig2}
\end{figure}

In Fig. \ref{fig2}, we show the size of a source as a function of the visibility 
amplitude in the longest baseline divided by that in the shortest baseline. The size shown 
in Fig. \ref{fig2} is given in units of the FWHM of the synthesized beam (we assume throughout 
this paper that {\em uniform weighting} is applied in the gridding of the visibilities, prior 
to the Fourier inversion; see Thomson, Moran, \& Swenson \cite{TMS} for more details). 
If the sensitivity of an interferometer allows us to detect a small decrease in the visibility 
amplitudes at the longest baselines, we are able to obtain information on the size of sources 
much smaller than the diffraction limit of 
the interferometer (i.e., much smaller than the FWHM of the PSF). The over-resolution 
power of an interferometer is, hence, dependent on the sensitivity of the observations, 
and can be {\em arbitrarily large}.

Figure \ref{fig2} has been computed using a different intensity profiles for the observed
source. It is obvious that the use of different source shapes (e.g., a 
Gaussian profile or a ring-like source) in a fit to the visibilities, results in different 
size estimates for the same dataset. 
Hence, if the structure of the observed source is similar to the model 
used in the fit to the visibilities (i.e., if we have a good a priori information on the real 
shape of the source), we can obtain precise estimates 
of sizes much smaller than the FWHM of the synthesized beam.

Indeed, the fact that the diffraction limit can be largely extrapolated by model fitting is 
well known since a long time, because in these cases the data are fitted with a simple 
model (i.e., with a small number of parameters), in contrast to the image-synthesis approach, 
where the super-resolution capabilities are much more limited due to the larger parameter 
space of the model (i.e., the image pixels). Therefore,
it is very difficult to obtain, from any data analysis based on the sky plane, size estimates of 
compact sources with a precision similar to that achieved in Fourier space. In the former case, 
the gridding of the images (i.e, the pixelation of the PSF), together with 
the particulars of the deconvolution algorithms, may smear out the fine details in the intensity 
profiles that encode the information on the structures of the underlying compact sources. On the 
other hand, compact sources on the sky are seen as very extended structures in Fourier 
space. Thus, a direct analysis of the visibilities (e.g., as described in Pearson \cite{Pearson})
is the optimum way to work with data coming from compact sources, since the effects of gridding 
in Fourier space will always be negligible.

\section{Maximum theoretical over-resolution power of an interferometer}
\label{MaxTheo}

As it is described in the previous section, the maximum over-resolution power of an 
interferometer (i.e., the minimum size of a source, in units of the FWHM of the synthesized 
beam, that can be resolved) depends on how precisely we can measure lower visibility 
amplitudes at the longest baselines. Hence, an interferometer with an arbitrarily
large sensitivity will accordingly have an arbitrarily large over-resolution power.

However, real interferometers have finite sensitivities, which depend on several 
factors (e.g., observing frequency, bandwidth, source coordinates, weather 
conditions,...). We can therefore ask the question of what is the minimum size 
of a source (relative to the diffraction limit of an interferometer) that still allows
us to extract a size information from the observed visibilities. 

In the extreme case of a very compact (and/or weak) source, such that it is not possible 
to estimate a statistically-significant lowering in the
visibility amplitudes at the longest baselines (because of the noise contribution to the 
visibilties), the only meaningful statistical analysis that can still be applied to the data 
is to estimate an {\em upper limit} to the source size by means of hypothesis testing.

Let us observe a source and assume the null hypothesis, $H_0$, that it 
has no structure at all (i.e., the source is point-like, so the visibility amplitudes are 
constant through the whole Fourier space). The likelihood, $\Lambda_0$, corresponding to a 
point-like model source fitted to the visibilities is

\begin{equation}
\Lambda_0 \propto \exp{\left( -\chi_0^2 \right)} = \exp{\left( - \sum_{j,k}^{N} { F^{jk}(V_j-S_0)(V_k-S_0) } \right)}
\label{chinull}
\end{equation}

\noindent where $N$ is the number of visibilities, $V_j = A_j\exp{(i\phi_j)}$ is the $j$-th 
visibility, $S_0$ is the maximum-likelihood (ML) estimate of the source flux density (it can 
be shown to be equal to the real part of the weighted visibility average, $\langle V \rangle$), 
and $F$ is the Fisher matrix of the visibilities (i.e., the inverse of the their 
covariance matrix), i.e.

$$ F = C^{-1} \mathrm{~~~with~~~} C_{j,k} = \rho_{jk} \sigma_j \sigma_k.$$

In this equation, $\rho_{jk}$ is the correlation coefficient between the $j$-th and 
$k$-th visibility, and $\sigma_j$ is the uncertainty of the $j$-th visibility.
Equation \ref{chinull} is generic and accounts for any correlation in the 
visibilities (e.g., possible global or antenna-dependent amplitude biases).
It can be shown that $\log{\Lambda_0}$ follows a $\chi^2$ distribution (close to 
its maximum) with a number of degrees of freedom equal to the 
rank of matrix $F$ minus unity. In the particular case when there is no correlation
between visibilities\footnote{Indeed, this also holds when there {\em is} correlation, 
unless in the pathological cases when the covariance matrix may be degenerate.}, 
the number of degrees of freedom equals $N-1$.

Let us model the visibilities with a function, $V^m(S,q\,\theta)$, that corresponds 
to the model of a (symmetric) source of size $\theta$ and flux density $S$ (the model 
amplitude, $V^m$, depends on the product of $\theta$ times the distance in Fourier space, 
$q$). Hence, the visibility $V_j$ is modelled by the amplitude $V^m_j = V^m(S,q_j\,\theta)$. 
The likelihood of this new modelling, $\Lambda_m$, is also given by 
Eq. \ref{chinull}, but changing $S_0$ by $V^m$. In addition, the distribution of 
$\log{\Lambda_m}$ (close to its maximum value) also follows a $\chi^2$ distribution, but
with one degree of freedom less than that of $\log{\Lambda_0}$.



Let us ask the question of what is the maximum value of $\theta$ (we call it 
$\theta_M$) corresponding to a value of the log-likelihood of $V^m$ that 
is compatible, by chance, with the parent distribution of the log-likelihood of a 
point source. We will estimate $\theta_M$ by computing a critical probability for the 
hypothesis that both quantities come from the same parent distribution (this is, indeed, 
our null hypothesis, $H_0$). Critical probabilities of 5\% and 0.3\% will be used in our h
ypothesis testing (these values correspond 
to the 2$\sigma$ and 3$\sigma$ cutoffs of a Gaussian distribution, respectively). The value of 
$\theta_M$ estimated in this way will be the maximum size that the observed source may have, 
such that the interferometer could have measured the observed visibilities with a chance given 
by the critical probability of $H_0$. 

The log-likelihood ratio (in our case, the difference of chi-squares; see Mood, Franklin, \& 
Duan \cite{Estat}) between the model of a point source and that of a source with visibilities 
modelled by $V^m$ 
follows a $\chi^2$ distribution with one degree of freedom, as long as $N$ is large 
(e.g., Wilks \cite{Wilks}). Hence, $H_0$ will not be discarded as long as 

\begin{equation}
\log{\Lambda_m} - \log{\Lambda_0} < \lambda_c/2,
\label{Ratio1}
\end{equation} 

\noindent where $\lambda_c$ is the value of the log-likelihood corresponding to the critical 
probability of the null hypothesis, assuming a $\chi^2$ distribution with one degree of freedom. 
$\lambda_c$ takes the values 3.84 or 8.81 (for a 5\% and a 0.3\% probability cuttoff of $H_0$, 
respectively). Working out Eq. \ref{Ratio1}, 
we arrive to




\begin{equation}
\sum_{jk}^N{\left((V^m_j - 2\,V_j)V^m_k+2\,S_0V_k -S_0^2 \right)F^{jk}} = \frac{\lambda_c}{2}.
\label{Carnati1}
\end{equation}

We can simplify Eq. \ref{Carnati1} in the special case when the off-diagonal elements in 
the covariance matrix are small (so the visibilities are nearly independent). On the one
hand, the standard deviation, $\sigma$, of the weighted visibility average is 
$ 1/\sigma^2 = \sum_j{F_{jj}} = \sum_{j}{1/\sigma_j^2}$; on the other hand, the weighted 
average of any visibility-related quantity, $A$, is 
$\langle A \rangle = \sigma^2 \sum_j{F_{jj}A_j}$. Hence, if $F_{jk} \sim 0$ for $j \neq k$
(i.e., if both matrices, $F$ and $C = F^{-1}$ are diagonal),
we have

\begin{equation}
\langle \left(V^m\right)^2 + S_0^2 - 2 V V^m \rangle \sim \langle \left(S_0 - V^m \right)^2 \rangle = \frac{\lambda_c \sigma^2}{2},
\label{Carnati1.5}
\end{equation}

\noindent i.e., the critical probability in our hypothesis testing depends on the 
weighted average of the quadratic difference between the two models being compared (a 
point source with flux density $ S_0 = \langle V \rangle$ and an extended source with 
a model given by $V^m$). In this equation, it is assumed that 
$\langle V V^m \rangle \sim \langle S_0 V^m \rangle$. Indeed,

$$\langle V V^m \rangle = \langle S_0 V^m \rangle + \langle (V-S_0) V^m \rangle. $$

Since the source is compact in terms of the diffraction limit of the interferometer 
(otherwise, this hypothesis testing would not be meaningful), the dependence of 
$V^m$ on the distance in Fourier space, $q$, will be small (i.e., the best-fitting 
model $V^m$ will almost be constant for the whole set of observations). 
As long as the decrease in $V^m$ with $q$ is smaller than the standard deviation of the 
visibilities (which holds if there is no hint of a decrease of visibility amplitudes 
with $q$), the quantity $\langle (V-S_0) V^m \rangle$ will always approach zero
(notice that $\langle (V-S_0) \rangle = 0$). 
Hence, from Eq. \ref{Carnati1.5}, we can finally write

\begin{equation}
\langle \left( 1 - \frac{V^m}{\langle V \rangle} \right)^2 \rangle = \frac{\lambda_c}{2(\mathrm{SNR})^2}
\label{Carnati2}
\end{equation}

\noindent where SNR is the signal-to-noise ratio of the weighted visibility 
average (i.e., $\langle V \rangle /\sigma$). If the covariance matrix is far from diagonal, 
Eq. \ref{Carnati2} will not apply, since the off-diagonal elements in Eq. \ref{Carnati1} 
would be added to the left-hand side of Eq. \ref{Carnati2}. 
However, it can be shown that if all the off-diagonal elements of $F_{jk}$ are roughly 
equal and $\langle V \rangle \sim \langle V^m \rangle \sim S_0$ (which is true if both 
models, $V^m$ and a point source $S_0$, satisfactorily fit to the data) then the combined 
effect of the off-diagonal elements in Eq. \ref{Carnati1} cancels out, and
Eq. \ref{Carnati2} still applies.

This restriction for the $F$ matrix (and hence for the correlation matrix) can be interpreted 
in the following way. A distribution of visibilities following a covariance matrix with 
roughly equal off-diagonal elements implies that 
all the antennas in the interferometer should have similar sensitivities, and the baselines 
related to each one of them should cover the full range of distances in Fourier space. 
If these conditions are fulfilled, it is always possible to remove any bias in the antenna
gains, and produce a set of visibilities with a roughly equal covariance between them. 
If, for instance, the gain at one antenna was biased, the visibilities of 
the baselines of all the other antennas would have 
different amplitudes in similar regions of the Fourier space, hence allowing us to correct 
for that bias by means of amplitude self-calibration. However, if the array was sparse, 
there might be antenna-related amplitude biases affecting visibilities at disjoint regions 
of Fourier space, thus preventing the correction for these biases using {\em closeby}
measurements from baselines of other antennas.
Interferometers with a sparse distribution of elements may thus produce sets of visibilities 
with different covariances (stronger at closer regions in 
Fourier space), thus making it difficult to estimate $\theta_M$ correctly (unless in the
very unlikely cases when the {\em whole} matrix $F$ is known!). This is 
the case of the NRAO Very Long Baseline Array (VLBA), where the antennas at Mauna Kea and 
St. Croix only appear in the longest baselines; lower visibility amplitudes at these 
baselines could be thus related to biased antenna gains, instead of source structure.



The new interferometric 
arrays, made of many similar elements with a smooth spatial distribution (as ALMA or the SKA) 
almost fulfill the condition of homogeneous Fourier coverage described here (i.e., there are 
almost no antennas exclusively appearing at long or short baselines), and are hence 
very robust for the over-resolution of compact sources well below 
their diffraction limits. 

Equation \ref{Carnati2} allows us to estimate the value of $\theta_M$ from a given 
distribution of baseline lengths, $q_j$, and for a given SNR in the weighted visibility average. 
Let us now assume that we have a very large number of visibilities 
and the sampling of baseline lengths, $q_j$, is quasi-continuous. We then have

\begin{equation}
\int_0^Q{n(q)\,(1-\frac{1}{\langle V \rangle}V_m(S,q\,\theta_M))^2\,dq} = \frac{\lambda_c}{2(\mathrm{SNR})^2},
\label{cond3}
\end{equation}

\noindent where $n(q)$ is the (normalized) density of visibilities at a distance $q$ in 
Fourier space 
%
%
%
%
and $Q$ is the maximum baseline length of 
the interferometer. 
Usually, $n(q)$ is large for small values of $q$ and decreases with increasing $q$ (i.e., the number 
of short baselines is usually larger than the number of long baselines)\footnote{This statement may 
not hold in special cases where the array consists of a few distant compact subarrays, whose elements
are considered as independent parts of the interferometer.}. 
The effect of $n(q)$ on $\theta_M$ is such that an interferometer with a large number of 
long baselines has a higher over-resolution power than another interferometer with a lower 
number of long baselines, {\em even} if the maximum baseline length, $Q$, is the same for 
both interferometers. The over-resolution power can also be increased if we decrease
the right-hand sides of Eqs. \ref{Carnati2} and \ref{cond3}. This can be achieved by increasing the 
sensitivity of the antennas and/or the observing time, {\em even} if the maximum baseline length 
is unchanged.

\begin{figure}
\centering
\includegraphics[width=8cm]{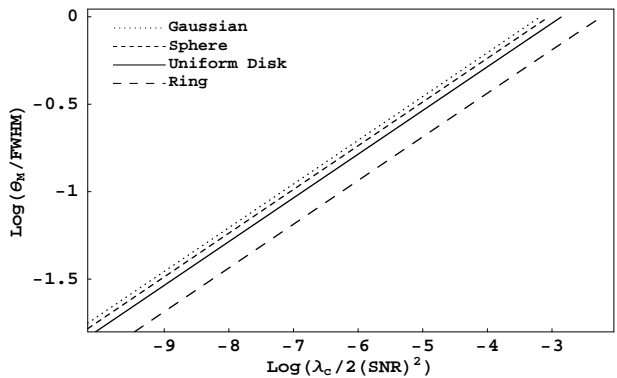}
\caption{Minimum detectable size of a source (in units of the FWHM of the synthesized 
beam) as a function of the source model (for a constant density of baseline lengths).}
\label{fig3}
\end{figure}

We show in Fig. \ref{fig3} the value of $\theta_M$ (in units of the FWHM of the synthesized 
beam) corresponding to different array sensitivities 
(i.e., different SNR in the visibility average) and source models (i.e., 
Gaussian, uniform-disk, sphere, and ring). 
He have used a baseline-length distribution with constant density (a constant $n(q)$). 
In all cases, the over-resolution power of the interferometer can be very well 
approximated by the following expression

\begin{equation}
 \theta_M = \beta\,\left( \frac{\lambda_c}{2(\mathrm{SNR})^2} \right)^{1/4} \times \mathrm{FWHM},
\label{PhenoEq}
\end{equation}

\noindent where $\beta$ slightly depends on the shape of $n(q)$ and the intensity profile of 
the source model. It usually takes values in the range 0.5--1.0 (it is larger for steeper $n(q)$ 
and/or for source intensity profiles with higher intensities at smaller scales). This 
equation is very similar to Eq. A.9 in Lobanov (\cite{Lobanov}), although we notice that
in the more general 
case, $\theta_M$ should be solved directly from Eq. \ref{Carnati2} (or even from Eq. \ref{Carnati1}, 
if the array was sparse and/or the covariance matrix was far from homogeneous or diagonal).
We notice that $\theta_M$ in Fig. \ref{fig3} can be interpreted in two different ways; either as 
the maximum possible size of a source that generates visibilities compatible with a point-like 
source or as the {\em true} minimum size of a source that can still be resolved by the 
interferometer. Any of these two (equivalent) interpretations of $\theta_M$ lead us to the 
conclusion that an interferometer is capable of resolving structures well below the mere 
diffraction limit achieved in the aperture synthesis.

In the cases of ultra-sensitive interferometric arrays like the SKA, where dynamic 
ranges of even $10^6$ will be eventually achieved in the images, 
the over-resolution power in observations 
of strong and compact sources can be very large. As an example, a dedicated 
observation with the SKA (let us assume 200 antennas) during one hour (with an integration 
time of 2 seconds), and an SNR of 100 for each visibility, results in a minimum resolvable 
size of only $\sim$$2\times10^{-3}$ times the FWHM of the synthesized beam (Eq. \ref{PhenoEq}). 
As a result, and depending on the 
observing frequency, the over-resolution power of the SKA would allow us to study details of 
sources at angular scales down to a few $\mu$as. This is, indeed, a resolution higher than 
the diffraction limit achieved with the current VLBI arrays (i.e., using much longer baselines).

\section{Summary}

We have reviewed the effects of source compactness in interferometric observations. 
The analysis of visibilities in Fourier space allows us to estimate sizes 
of very compact sources (much smaller than the diffraction limit achieved in the aperture
synthesis). As the sensitivity 
of the interferometer increases, the minimum size of the sources that can still 
be resolved decreases (i.e., the over-resolution power of the interferometer increases). In this 
sense, the analysis of observations of very compact sources in Fourier space is 
more reliable (and robust) than alternative analyses based on synthesized 
images of the sky intensity distribution (and affected by beam gridding, deconvolution biases, etc.). 

We study the case of extremely compact sources observed with an intereferometer of finite
sensitivity. If the source is such compact and/or weak that it is not possible to detect 
structure in the visibilities, we describe a test of hypothesis to set a strong 
upper limit to the size of the source.
We also compute the minimum possible size of a source whose structure can still be 
resolved by an interferometer (i.e., the maximum theoretical over-resolution power of an 
interferometer, computed from Eq. \ref{Carnati2} and approximated in Eq. \ref{PhenoEq}). 
The over-resolution power depends on the number of visibilities, the array sensitivity, 
and the spatial distribution of the baselines, and increases if 1) the 
number of long baselines increases (i.e., not necessarily the {\em maximum} baseline 
length, but only the {\em number} of long baselines relative to the number of short 
baselines); 2) the observing time increases; and/or 3) the array sensitivity increases. 

\begin{acknowledgements}

The authors are thankful to J.M. Anderson for discussion and to the anonymous
referee for his/her useful comments.
MAPT acknowledges support through grant AYA2006-14986-C02-01 (MEC), and 
grants FQM-1747 and TIC-126 (CICE, Junta de Andaluc\'ia).

\end{acknowledgements}


\begin{thebibliography}{.99}


\bibitem[2005]{Kovalev} Kovalev Y.Y., Kellerman K.I., Lister, M.L., et al. 2005, AJ, 130, 2473

\bibitem[2005]{Lobanov} Lobanov, A.P. 2005, arXiv:astro-ph/0503225

\bibitem[1974]{Estat} Mood, A., Franklin A.G., \& Duane, C.B. 1974, 
Introduction to the Theory of Statistics, McGraw-Hill, New York

\bibitem[1999]{Pearson} Pearson, T.J. 1999, ASPC, 180, 335

\bibitem[1986]{TMS} Thomson A.R., Moran, J.M. \& Swenson, G.W. 1986, Interferometry 
and Synthesis in Radio Astronomy, Wiley, New York

\bibitem[1938]{Wilks} Wilks, S.S. 1938, Ann. Math. Statist., 9, 1 

\end{thebibliography}
\end{document}